\documentclass[floatfix,aps,prd,showpacs,amsmath,nofootinbib,preprintnumbers,twocolumn]{revtex4}

\usepackage{graphicx}
\usepackage{bm}

\newcommand{\figurewidths}{4.2in}
\newcommand{\figurewidthw}{5.2in}

\def\half{{1\over 2}}

\def\half{{1\over 2}}
\def\({\left(}
\def\){\right)}
\def\[{\left[}
\def\]{\right]}

\def\e{\begin{equation}}
\def\q{\end{equation}}
\def\m{\begin{eqnarray}}
\def\n{\end{eqnarray}}

\begin{document}

\title{Constraints on the extensions to the base $\Lambda$CDM model from BICEP2, Planck and WMAP}

\author{Cheng Cheng$^{1,2}$ \footnote{chcheng@itp.ac.cn}, Qing-Guo Huang$^1$ \footnote{huangqg@itp.ac.cn} and Wen Zhao$^3$ \footnote{wzhao7@ustc.edu.cn}}
\affiliation{$^1$ State Key Laboratory of Theoretical Physics,
Institute of Theory Physics, Chinese Academy of Science, Beijing
100190, China}
\affiliation{$^2$ University of the Chinese Academy of Sciences, Beijing 100190, China}
\affiliation{$^3$ Key Laboratory for Researches in Galaxies and Cosmology,
Department of Astronomy, University of Science and Technology of China, Hefei, Anhui,
230026, China}

\date{\today}

\begin{abstract}

Recently Background Imaging of Cosmic Extragalactic Polarization
(B2) discovered the relic gravitational waves at $7.0\sigma$
confidence level. However, the other cosmic microwave background
(CMB) data, for example Planck data released in 2013 (P13), prefer
a much smaller amplitude of the primordial gravitational waves
spectrum if a power-law spectrum of adiabatic scalar perturbations
is assumed in the six-parameter $\Lambda$CDM cosmology. In this
paper, we explore whether the $w$CDM model and the running
spectral index can relax the tension between B2 and other CMB
data. In particular, we find that a positive running of running of
spectral index is preferred at $1.7\sigma$ level from the
combination of B2, P13 and WMAP Polarization data.

\end{abstract}


\maketitle

%%%%%%%%%%%%%%%%%%%%%%%%%%%%%%%%%%%%%%%%
%%%%%%%%%%%%%%%%%%%%%%%%%%%%%%%%%%%%%%%

\section{Introduction}

In the early of 2013, Planck (P13) \cite{Ade:2013zuv} released its data
which precisely measured the temperature anisotropies of cosmic
microwave background (CMB), and claimed that it strongly supports
the standard spatially-flat six-parameter $\Lambda$CDM cosmology
with a power-law spectrum of adiabatic scalar perturbations.
Actually the relic gravitational waves could also make
contributions to the temperature and polarization power spectra in
the CMB \cite{Grishchuk:1974ny,Starobinsky:1979ty,Rubakov:1982,Crittenden:1993ni,Krauss:2013pha}. Combining Wilkinson Microwave Anisotropy Probe (WMAP) 9-year data \cite{Hinshaw:2012aka} with Baryon Acoustic Oscillation (BAO) \cite{BAO}, $H_0$ prior from Hubble Space Telescope (HST) \cite{Riess:2011yx} and other highL CMB data,
including Atacama Cosmology Telescope (ACT) \cite{Sievers:2013ica}
and South Pole Telescope (SPT) \cite{Story:2012wx}, we obtained
the constraint on the primordial gravitational waves before Planck
as follows
\m
r_{0.002}<0.12
\label{rw9}
\n
at $95\%$ CL in \cite{Cheng:2013iya}, where $r_{0.002}$ is the tensor-to-scalar
ratio at the pivot scale $k_p=0.002$ Mpc$^{-1}$ and a power-law
spectrum of the primordial scalar perturbations is also assumed. A similar result was reported by Planck combining with WMAP polarization (WP) data and other highL CMB data, namely
\m
r_{0.002}<0.11
\label{rp13}
\n
at $95\%$ CL in \cite{Ade:2013zuv}. In this paper, we shall fix the
pivot scale as $k_p=0.002$ Mpc$^{-1}$.

Considering that the primordial gravitational waves only make
contributions to CMB power spectra at the very large scales, we
fixed the background parameters as their best-fit values from
Planck, and then run the CosmoMC to figure out the amplitude of
adiabatic scalar perturbations, spectral index and the
tensor-to-scalar ratio by only using the low-multipole Planck TT \cite{Ade:2013zuv} and
WMAP TE (WP) data \cite{Hinshaw:2012aka}. We found $r>0$ at more than 68$\%$ confidence
level with maximum likelihood at around $r\sim 0.2$ \cite{Zhao:2014rna}. Our new result confirmed the
previous one in \cite{Zhao:2010ic} where WMAP 7-year data were
utilized. Recently Background Imaging of Cosmic
Extragalactic Polarization (B2) \cite{Ade:2014xna} discovered the relic gravitational waves with the tensor-to-scalar ratio
\m
r=0.20_{-0.05}^{+0.07},
\label{rb14}
\n
and $r=0$ is disfavored at $7.0\sigma$. Using B2 only or the combination of B2, P13 and WP, the tilt $n_t$ of relic gravitational waves spectrum is constrained to be around zero and $n_t=2$ is ruled out at more than $5\sigma$ confidence level in \cite{Cheng:2014bma,Cheng:2014ota} which strongly indicates that inflation \cite{Guth:1980zm,Linde:1981mu,Albrecht:1982wi} really happened in the early Universe.

In this paper we hope to get a better understanding about the
physics in our Universe through a more careful investigation of
the datasets. Comparing (\ref{rb14}) to (\ref{rw9}) and
(\ref{rp13}), we see that there is a moderately strong tension
between B2 and other CMB data in the base six-parameter
$\Lambda$CDM+tensor cosmology. If all of these CMB datasets are
trustable, it strongly implies that our Universe is much more
complicated than what we expected before. In order to reconcile
the tension on constraining the primordial gravitational waves
between P13 and B2, we need to go beyond the $\Lambda$CDM+tensor
model.
There are several well-motivated extensions to the $\Lambda$CDM+tensor model which might relax such an inconsistency.  \\
i) More complicated physics in the early
universe can be involved. Here we consider that the spectrum of
adiabatic scalar perturbations departures from a pure power-law
form, and the running of spectral index ($dn_s/d\ln k$) and the running of running ($d^2 n_s/d\ln k^2$) are taken into
account. Or the spatial curvature ($\Omega_k$) of our Universe
deviates from exact flatness. \\
ii) We can consider more
complicated physics about neutrino and relativistic components by
relaxing the total mass of active neutrinos ($\sum m_\nu$), or the
number of relativistic species ($N_{\rm eff}$). \\
iii) The abundance of light elements, for example $Y_P\equiv 4n_{\rm
He}/n_{\rm b}$ for helium-4, is taken as a free parameter. \\
iv) The dark energy is not a cosmological constant and its
equation-of-state (EOS) parameter $w\equiv p_{\rm de}/\rho_{\rm
de}$ is regarded as
a free parameter.\\
An almost comprehensive investigation has been given by Lewis in
\cite{lewis:2014d} where the combination of B2 and P13 was
considered. In this paper we will adopt B2, P13 and WP to explore
two extensions: one is to relax the dark energy model from
cosmological constant to one with a constant EOS parameter
$w=p_{\rm de}/\rho_{\rm de}$; the other is to take into account
the running of spectral index and the running of running. Here we
fix the consistency relation to be $n_t=-r/8$.

%Our paper is organized as follows. In the next section, our main
%results will be presented. Conclusion and discussion are given in
%section III.

\section{Implications for cosmology from BICEP, Planck and WMAP}

In this section we will use the CosmoMC \cite{cosmomc} to work out the constraints on the
cosmological parameters in different cosmological models from
several different combinations of datasets respectively. Our results are summarized in
Tables~\ref{tab:w}, \ref{tab:nrun} and \ref{tab:nrunrun}.

\subsection{$w$CDM model}

In this subsection, we extend the dark energy model from the
cosmological constant to the dark energy with constant EOS
parameter $w$. As we know, there are also several tensions between
P13 and some local cosmological observations, including the $H_0$
prior from HST \cite{Riess:2011yx} and Supernova Legacy Survey
(SNLS) samples \cite{Conley:2011ku}. For example, P13 prefers a
larger matter density today compared to SNLS and a smaller Hubble
constant compared to the $H_0$ prior from HST. However these two
tensions can be significantly relaxed in the $w$CDM model
\cite{Cheng:2013csa} where the dark energy is preferred to be
phantom-like, namely $w=-1.16\pm 0.06$ from the combination of
P13+WP+BAO+SNLS+HST.

\begin{figure*}[hts]
\begin{center}
\includegraphics[width=\figurewidths]{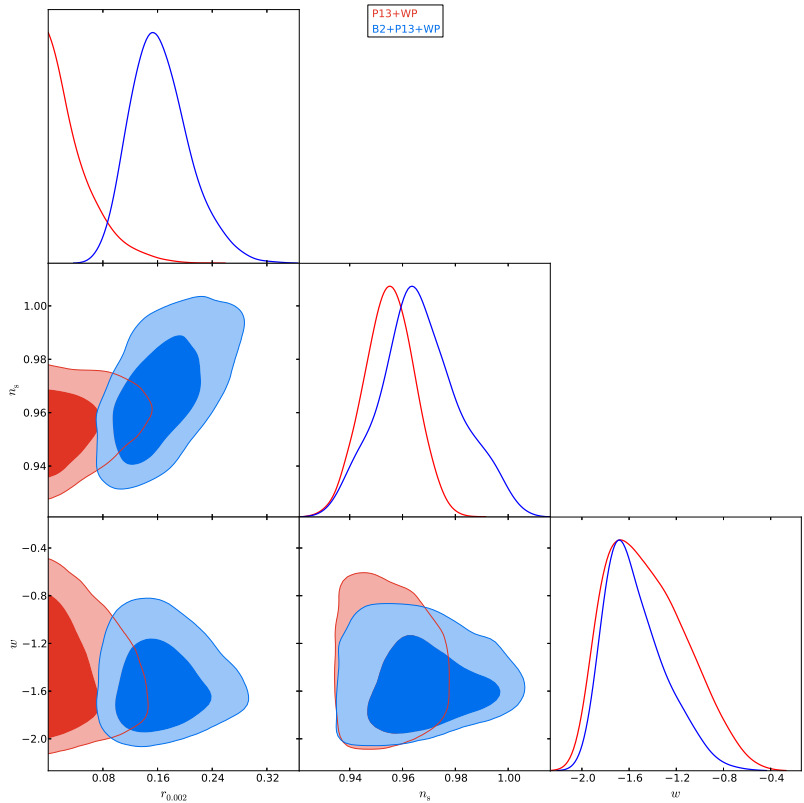}
\end{center}
\caption{The constraint contours on $r$, $n_s$ and $w$ from the
combinations of P13+WP and B2+P13+WP in the $w$CDM+r model. }
\label{fig:w}
\end{figure*}

Here we also wonder whether the dark energy EOS can help to relax
the tension on the tensor-to-scalar ratio between P13 and B2. We
constrain the cosmological parameters in the $w$CDM+r model by
adopting the combinations of P13+WP and B2+P13+WP, respectively.
See the results in Table \ref{tab:w} and Fig.~\ref{fig:w}.
\begin{table}[htbp]
\centering
\renewcommand{\arraystretch}{1.5}
\scriptsize
{

\

\begin{tabular}{c|cc}
\hline\hline
$w$CDM+r & \multicolumn{2}{|c}{B2+P13+WP} \\
\hline
parameters & best fit &$68\%$ limits   \\
\hline
$\Omega_bh^2$ & 0.02222 & $0.02210_{-0.00063}^{+0.00058}$ \\
$\Omega_ch^2$ & 0.1161 & $0.1172_{-0.0028}^{+0.0029}$ \\
100$\theta_{\rm MC}$ & 1.04190 & $1.04163_{-0.00068}^{+0.00067}$  \\
$\tau$ & 0.1001 & $0.0888_{-0.0186}^{+0.0145}$ \\
$\ln (10^{10}A_s)$ & 3.200 & $3.186\pm 0.034$ \\
$n_s$ & 0.9690 & $0.9669_{-0.0163}^{+0.0140}$ \\
$r_{0.002}$& 0.16 & $0.16_{-0.05}^{+0.04}$  \\
$w$& -1.70  & $-1.54_{-0.32}^{+0.17}$  \\
\hline
\end{tabular}
}
\caption{Constraints on the cosmological parameters in the $w$CDM+r model.  }
\label{tab:w}
\end{table}
%%%%%%%%%%%%%%%%%%%%%%%
We find that the constraint on $r$ is given by \m r_{0.002}<0.16
\n at $3\sigma$ confidence level from P13+WP. There is still a more than $3\sigma$ tension on $r$
between P13+WP and B2 in the $w$CDM+r model. Therefore relaxing
the dark energy model cannot reconcile the tension on $r$ between
P13 and B2. Due to such a
tension, some exotic results appear. For example, the constraint
on the dark energy EOS parameter becomes $w=-1.54_{-0.32}^{+0.17}$
in Table.~\ref{tab:w}. A similar constraint on $w$ from B2+P13 is
$w=-1.55_{-0.34}^{+0.18}$ in \cite{Li:2014cka}.

\subsection{Running spectral index}

In this subsection, we extend the six-parameter base
$\Lambda$CDM+r model to the $\Lambda$CDM+nrun+r and
$\Lambda$CDM+nrun+nrunrun+r models respectively, where nrun and
nrunrun denote the running of spectral index ($dn_s/d\ln k$) and
the running of running ($d^2 n_s/d\ln k^2$). In this
case the amplitude of scalar perturbation spectrum is parameterized by  \m
P_s(k)=A_s \({k\over k_p}\)^{n_s-1+\half {dn_s\over d\ln k}\ln
{k\over k_p}+{1\over 6} {d^2n_s\over d\ln k^2} \ln^2 {k\over
k_p}}. \n In \cite{Cheng:2013iya}, the constraint on the
tensor-to-scalar ratio from the combination of
WMAP+ACT+SPT+BAO+HST is relaxed to be \m r_{0.002}<0.42 \n at
$95\%$ CL if the running of spectral index is considered, and \m
r_{0.002}<0.53 \n at $95\%$ CL if both the running and running of
running are taken into account. In \cite{Ade:2013zuv}, the
constraint on the tensor-to-scalar ratio is relaxed to be \m
r_{0.002}<0.26 \n at $95\%$ CL from the combination of
P13+WP+ACT+SPT if the running of spectral index is considered. We see that the constraint on $r$ can be significantly
loosen to be consistent with B2 in the model with running spectral
index.

\begin{figure*}[hts]
\begin{center}
\includegraphics[width=\figurewidths]{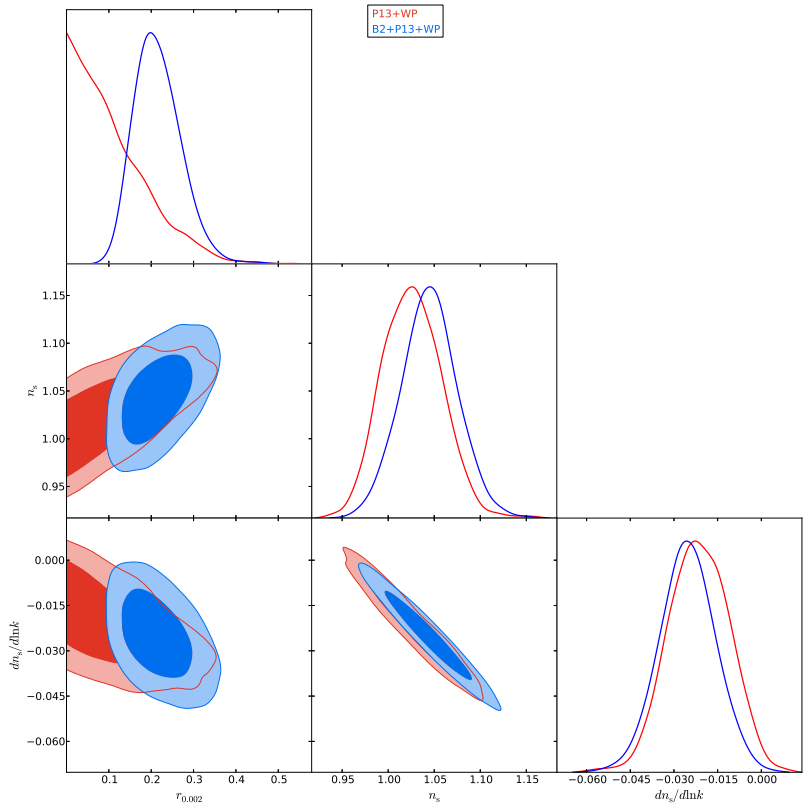}
\end{center}
\caption{The constraint contours on $r$, $n_s$ and $dn_s/d\ln k$ from the combinations of P13+WP and B2+P13+WP in the $\Lambda$CDM+nrun+r model.
}
\label{fig:nrun}
\end{figure*}

First of all, we combine B2 with P13 and WP to constrain the cosmological parameters in the $\Lambda$CDM+nrun+r cosmology. Our results are given in Table \ref{tab:nrun} and Fig.~\ref{fig:nrun}.
\begin{table}[htbp]
\centering
\renewcommand{\arraystretch}{1.5}
\scriptsize
{

\

\begin{tabular}{c|cc}
\hline\hline
$\Lambda$CDM+nrun+r & \multicolumn{2}{|c}{B2+P13+WP}  \\
\hline
parameters&Best fit  &$68\%$ limits  \\
\hline
$\Omega_b h^2$ & 0.02229 & $0.02246_{-0.00032}^{+0.00030}$ \\
$\Omega_c h^2$ & 0.1187 & $0.1173_{-0.0021}^{+0.0022}$ \\
100$\theta_{\rm MC}$ & 1.04154 & $1.04162_{-0.00062}^{+0.00061}$ \\
$\tau$ & 0.0991 & $0.1011_{-0.0161}^{+0.0137}$ \\
$\ln(10^{10}A_s)$ & 3.117 & $3.098_{-0.041}^{+0.045}$ \\
$n_s$ & 1.0336 & $1.0447_{-0.0297}^{+0.0295}$ \\
$dn_s/d\ln k$ & -0.0228 & $-0.0253\pm 0.0093$ \\
$r_{0.002}$& 0.18 & $0.22_{-0.07}^{+0.04}$  \\
\hline
\end{tabular}
}
\caption{Constraints on the cosmological parameters in the $\Lambda$CDM+nrun+tensor model.  }
\label{tab:nrun}
\end{table}
%%%%%%%%%%%%%%%%%%%%%%%%%%%%%%%%
We see that a blue tilted scalar power spectrum at $k_p=0.002$ Mpc$^{-1}$ is preferred at $1.5\sigma$ level, and a negative running of spectral index is favored at around $2.7\sigma$ level. Combining with P13, WP and other highL CMB data, B2 implies $dn_s/d\ln k=-0.028\pm 0.009$ \cite{Ade:2014xna}. In \cite{lewis:2014d}, the combination of B2+P13 gives a constraint $dn_s/d\ln k=-0.028\pm 0.020$. In \cite{Li:2014cka}, $dn_s/d\ln k=-0.0281\pm 0.0099$ from B2+P13+BAO+SN. See the analysis in \cite{Aslanyan:2014mqa,Abazajian:2014tqa} as well. Our results are consistent with all of these previous results.

\begin{figure*}[hts]
\begin{center}
\includegraphics[width=\figurewidthw]{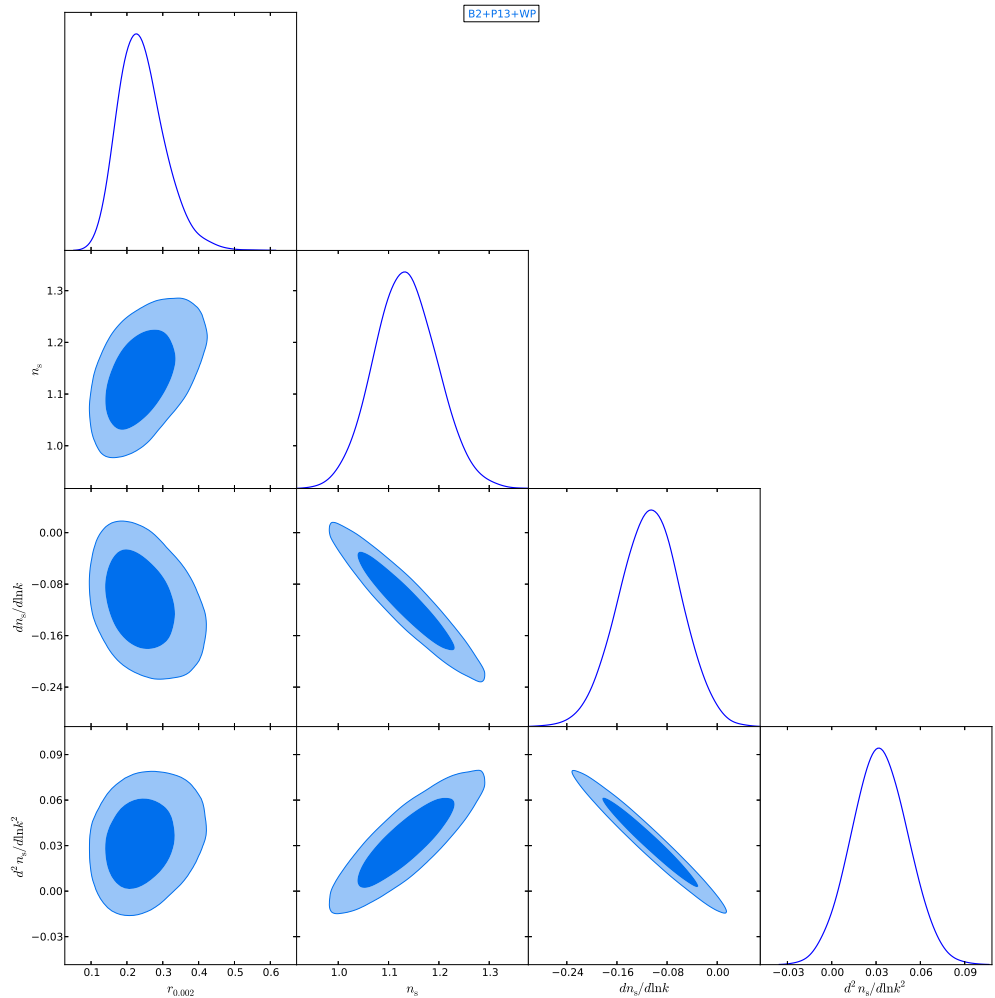}
\end{center}
\caption{The constraint contours on $r$, $n_s$, $dn_s/d\ln k$ and $d^2n_s/d\ln k^2$ from the combination of B2+P13+WP in the $\Lambda$CDM+nrun+nrunrun+r model.
}
\label{fig:nrunrun}
\end{figure*}

Since a negative running of spectral index is preferred at high confidence level, we wonder whether the higher order terms in the parametrization of scalar perturbation spectrum are required. Here we further extend the previous model to the $\Lambda$CDM+nrun+nrunrun+r model. The results show up in Table.~\ref{tab:nrunrun} and Fig.~\ref{fig:nrunrun}.
\begin{table}[htbp]
\centering
\renewcommand{\arraystretch}{1.5}
\scriptsize
{

\

\begin{tabular}{c|cc}
\hline\hline
$\Lambda$CDM+nrun+nrunrun+r & \multicolumn{2}{|c}{B2+P13+WP}  \\
\hline
parameters&Best fit  &$68\%$ limits  \\
\hline
$\Omega_b h^2$ & 0.02221 & $0.02217\pm{0.00035}$ \\
$\Omega_c h^2$ & 0.1203 & $0.1184\pm{0.0023}$ \\
100$\theta_{\rm MC}$ & 1.04145 & $1.04142_{-0.00063}^{+0.00064}$ \\
$\tau$ & 0.0983 & $0.1054_{-0.0168}^{+0.0142}$ \\
$\ln(10^{10}A_s)$ & 3.047 & $3.063_{-0.050}^{+0.066}$ \\
$n_s$ & 1.1656 & $1.1344_{-0.0608}^{+0.0612}$ \\
$dn_s/d\ln k$ & -0.139 & $-0.108_{-0.048}^{+ 0.049}$ \\
$d^2n_s/d\ln k^2$ & 0.045 & $0.033_{-0.019}^{+0.018}$ \\
$r_{0.002}$& 0.22 & $0.24_{-0.07}^{+0.05}$  \\
\hline
\end{tabular}
}
\caption{Constraints on the cosmological parameters in the $\Lambda$CDM+nrun+nrunrun+tensor model.  }
\label{tab:nrunrun}
\end{table}
%%%%%%%%%%%%%%%%%%%%%%%%%%%%%%%%
Compared to the previous model with only the running of spectral index, $\Delta\chi^2=9852.70-9855.82=-3.12$ which indicates that this further parameter extension is favored at more than $1\sigma$ level.
From Table \ref{tab:nrunrun}, we see that at the pivot scale $k_p=0.002$ Mpc$^{-1}$ the spectral index $n_s>1$ is preferred at $2.2\sigma$ level, a negative running of spectral index is preferred at $2.2\sigma$ level and a positive running of running is preferred at $1.7\sigma$ level once the running of running is considered. Our results imply that higher order expansions might be considered in the future as well.

\section{Summary and Discussion}

Discovery of relic gravitational waves opens a new window to explore cosmology. There are many possible sources for the relic gravitational waves, such as inflation \cite{Guth:1980zm,Linde:1981mu,Albrecht:1982wi}, cosmic string \cite{Pogosian:2003mz,Wyman:2005tu} and so on. In this paper we extend the $\Lambda$CDM+r cosmology to $w$CDM+r model and $\Lambda$CDM+r model with running spectral index, and find that the tension between B2 and P13 can be reconciled if a running spectral index is taken into account, but relaxing dark energy model does not work.

Usually inflation model predicts $|n_s-1|\lesssim {\cal
O}(10^{-2})$, $|dn_s/d\ln k|\lesssim {\cal O}(10^{-3})$ and
$|d^2n_s/d\ln k^2|\lesssim {\cal O}(10^{-4})$. Our results imply
that the simple canonical single-field slow-roll inflation models
are not compatible with the datasets and the physics in the early
Universe should be much more complicated than what we expect if
all of B2, P13 and WP are trustable. After B2 released its data,
many authors investigated inflation models widely. See, for
example,
\cite{Joergensen:2014rya,Nakayama:2014koa,Hamada:2014iga,Freese:2014nla,Gong:2014cqa,Okada:2014lxa,Bamba:2014jia,Lyth:2014yya,DiBari:2014oja,Feng:2014yja,Chung:2014woa}.
However almost all of them only tried to fit the value of
tensor-to-scalar ratio and the spectral index. We believe that it
is not enough because the combination of B2+P13+WP strongly
implies a running spectral index. How to natrually achieve a
significantly running spectral index is still an open question. As we known, the space-time non-commutative inflation \cite{Huang:2003zp,Huang:2003hw,Huang:2003fw} can generate a large negative running of spectral index. It can also be realized in the inflation with modulations \cite{Kobayashi:2010pz,Czerny:2014wua,Czerny:2014qqa} as well. 
Another possibility is that the Planck data is not reliable at
all. In \cite{Cheng:2014cja} we combine B2 with WMAP 9-year data
and find that the power-law spectrum of scalar perturbation is
compatible with B2+WMAP, and the power-law inflation and inflation
model with inverse power-law potential can fit the data nicely. In
a word, we believe that the realistic inflation model is still
unknown and further investigation is needed in the near future.

\vspace{5mm}
\noindent {\bf Acknowledgments}
We acknowledge the use of Planck Legacy Archive, ITP and Lenovo
Shenteng 7000 supercomputer in the Supercomputing Center of CAS
for providing computing resources. QGH is supported by NSFC
No.10821504, 11322545, 11335012 and project of KIP of CAS. WZ is
supported by project 973 under Grant No.2012CB821804, by NSFC
No.11173021, 11322324 and project of KIP of CAS.

%%%%%%%%%%%%%%%%%%%%%%%%%%%%%%%%%%%%%%%%
%%%%%%%%%%%%%%%%%%%%%%%%%%%%%%%%%%%%%%%%

%%%%%%%%%%%%%%%%%%%%%%%%%%%%%%%%%%%%%%%%
%%%%%%%%%%%%%%%%%%%%%%%%%%%%%%%%%%%%%%%%

%%%%%%%%%%%%%%%%%%%%%%%%%%%%%%%%%%%%%%%%
%%%%%%%%%%%%%%%%%%%%%%%%%%%%%%%%%%%%%%%%
\end{document}